\documentclass[12pt]{article}
\usepackage{amssymb,amsmath,bm}
\usepackage{fullpage}
\usepackage{mathtools}
\usepackage{booktabs}
\usepackage{algorithm2e}
\usepackage{tikz}
\usepackage{pgfplots}
\usepackage{natbib}

\usepackage[colorlinks=true,linkcolor=black,citecolor=black,urlcolor=blue,pdfborder={0 0 0}]{hyperref}

\newtheorem{theorem}{Theorem}

\begin{document}
\title{Guaranteed $\varepsilon$-optimal solutions with the linear optimizer ART3+O}
\author{Bram L. Gorissen}
\date{}
\maketitle
\abstract{The linear optimization algorithm ART3+O introduced by Chen et al.~(2010) can efficiently solve large scale inverse planning problems encountered in radiation therapy by iterative projection. Its major weakness is that it cannot guarantee $\varepsilon$-optimality of the final solution due to an arbitrary stopping criterion. We propose an improvement to ART3+O where the stopping criterion is based on Farkas' lemma. The same theory can be used to detect inconsistency in other projection methods as well. The proposed algorithm guarantees to find an $\varepsilon$-optimal solution in finite time. The algorithm is demonstrated on numerical examples in radiation therapy.}
\begin{tikzpicture}[remember picture,overlay]
\node[anchor=south,yshift=10pt] at (current page.south) {\fbox{\parbox{\dimexpr\textwidth-\fboxsep-\fboxrule\relax}{\footnotesize This is an author-created, un-copyedited version of an article published in Physics in Medicine and Biology \href{https://doi.org/10.1088/1361-6560/ab0a2e}{DOI:10.1088/1361-6560/ab0a2e}.}}};
\end{tikzpicture}
\section{Introduction}
A linear optimization problem is a linear objective subject to linear constraints:
\begin{align}\label{eq:standardform}
\max_x\{c^Tx : Ax\leq b, x \geq 0\}.
\end{align}
It is easy to show that any linear optimization problem can be written in this form, even if it is originally a minimization problem, has equality constraints or allows variables to be negative.

One application of linear optimization is inverse planning in radiation therapy, where the goal is to customize a treatment plan to the patient's anatomy. A 3d grid is superimposed onto a patient for discretization purposes. The cubes of this grid (voxels) that lie in or near the tumor (the planning target volume, PTV) get prescribed a certain dose level, while the other voxels, especially those in critical organs, should get as little dose as possible. In many treatment modalities, the dose in a voxel is linear in the fluence vector $x \in \mathbb{R}^n$ \citep{bortfeld1990methods,lessard2001inverse,lomax1999intensity}:
\begin{align}\label{eq:dose}
d_i = \sum_j D_{ij} x_j.
\end{align}
The coefficients $D_{ij}$ depend on the patient's anatomy and can be computed analytically or with monte carlo simulation. Surprisingly many dose objectives and constraints can be expressed in the framework of linear optimization. Mean and maximum dose \citep{bahr1968method}, mean dose above or under a given threshold dose \citep{shepard1999optimizing}, and mean dose in the tail \citep{romeijn2003novel} can be combined into the same optimization problem. For example, linear optimization can be used to find a treatment plan that minimizes the mean dose in the patient such that the mean dose below 50 Gy in the PTV is no more than 1 Gy, while the mean dose in the 5\% region of the PTV with the most dose does not exceed 55 Gy.

Although linear optimization is a versatile modeling tool, the solution time depends greatly on the structure of the problem. Lack of sparsity and the existence of many near-optimal solutions affect the solution time of off-the-shelf solvers to the level that they are not usable for clinical purposes. This has led to the development of new algorithms or more efficient implementations for linear optimization \citep{Breedveld2016,herman2008fast}, and the use of nonlinear optimization by many vendors of treatment planning systems \citep{trofimov2012treatment}.

One of the algorithms for linear optimization of radiation therapy is ART3+O \citep{herman2008fast,chen2010fast,chen2012including}. This algorithm is based on the algorithms ART3 by \citet{herman1975relaxation} that was originally proposed for image reconstruction, and its successor ART3+ \citep{herman2008fast}. ART3 and ART3+ fall in the category of projection algorithms, which have a long history \citep{bauschke1996projection}. Alg.~\ref{alg:art3} shows the outline of ART3, slightly simplified for our needs. The algorithm takes as input a matrix $A$, a vector $b$, and iteratively changes an arbitrary starting point $x$ until $Ax \leq b$. In iteration $k$, it considers the $i^{th}$ row of $A$, where $i=1,2,\ldots,m,1,2,\ldots,m,1,2,$ etc. If the $i^{th}$ inequality is violated, $x$ is reflected into the hyperplane $A_i x = b$. Figure \ref{fig:art3} depicts three iterations of ART3, and shows that a feasible point is found after two reflections. If the interior of the feasible set is full dimensional, ART3 finds a feasible point in a finite number of steps (vs in the limit) \citep{herman1975relaxation}. The key advantages of ART3 are its low memory requirements, fast practical running time, and simplicity.

The performance of ART3 has been improved by temporarily removing nonviolated constraints from the inner loop (Alg.~\ref{alg:art3plus}). This improved algorithm, ART3+, benefits from not having to check for a violation for those constraints that are not likely to be violated, which yields a significant speed-up \citep{herman2008fast}. Similar to ART3, ART3+ terminates in a finite number of steps if the feasible set is full dimensional.

\begin{algorithm}\label{alg:art3}
	\KwData{$A \in \mathbb{R}^{m \times n}$, $b \in \mathbb{R}^{m}$}
	\KwResult{$x$ such that $Ax \leq b$}
	Let the starting point $x \in \mathbb{R}^n$ be an arbitrary vector\;
	$0 \to k$\;
	\While{$Ax \nleq b$}{
		$1 + (k \mod m) \to i$\;
		$A_i x - b_i \to v_i$\;
		\If{$v_i > 0$}{
			$x - 2\frac{v_i}{||A_i||^2} A_i^T \to x$
		}
		$k + 1 \to k$\;
	}
\caption{The image reconstruction algorithm ART3 \citep{herman1975relaxation}.}
\end{algorithm}
\begin{algorithm}\label{alg:art3plus}
	\KwData{$A \in \mathbb{R}^{m \times n}$, $b \in \mathbb{R}^{m}$}
	\KwResult{$x$ such that $Ax \leq b$}
	Let the starting point $x \in \mathbb{R}^n$ be an arbitrary vector\;
	\While{$Ax \nleq b$}{
		$\{1,2,\ldots,m\} \to S$\;
        \While{$S \neq \emptyset$}{
			Let $i$ be the next element in $S$\;
			$A_i x - b_i \to v_i$\;
			\eIf{$v_i > 0$}{
				$x - 2\frac{v_i}{||A_i||^2} A_i^T \to x$
			}{$S\backslash\{i\} \to S$}
		}
	}
	\caption{The algorithm ART3+ \citep{herman2008fast}.}
\end{algorithm}
\begin{algorithm}\label{alg:art3plusO}
	\KwData{$A \in \mathbb{R}^{m \times n}$, $b \in \mathbb{R}^{m}$, $c \in \mathbb{R}^n$, $L \in \mathbb{R}$, $U \in \mathbb{R}$}
	\KwResult{$x$ that solves \eqref{eq:standardform}}
	\While{$U-L > \varepsilon$}{
		$(L+U)/2 \to M$\;
		Call ART3+ to find $x$ such that $\begin{pmatrix}-c^T \\ A \\ -I\end{pmatrix} x \leq \begin{pmatrix} -M \\ b \\ 0\end{pmatrix}$\;
		\eIf{ART3+ found such an $x$}{$c^Tx \to L$}{$M \to U$}
	}
	\caption{The algorithm ART3+O \citep{chen2010fast}.}
\end{algorithm}
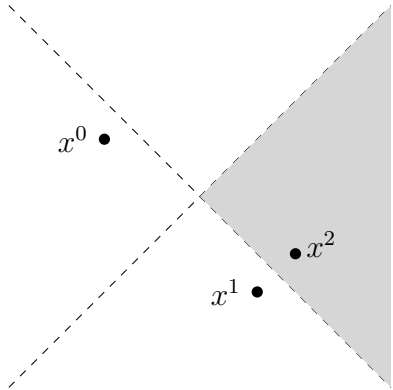
\begin{figure}\label{fig:art3}
\centering
%
%
\begin{tikzpicture}

\begin{axis}[%
width=2in,
height=2in,
at={(0.758in,0.481in)},
scale only axis,
xmin=-2,
xmax=2,
ymin=-2,
ymax=2,
axis line style={draw=none},
ticks=none
]
\addplot [color=black, dashed, forget plot]
  table[row sep=crcr]{%
-2	-2\\
2	2\\
};
\addplot [color=black, dashed, forget plot]
  table[row sep=crcr]{%
-2	2\\
2	-2\\
};

\addplot[area legend, draw=none, fill=white!84!black, forget plot]
table[row sep=crcr] {%
x	y\\
0	0\\
2	2\\
2	-2\\
}--cycle;
\node at (axis cs:-1,0.6)[circle,fill,inner sep=1.5pt]{};
\node[right, align=left] at (axis cs:-1.6,0.6) {$x^0$};

\node at (axis cs:0.6,-1)[circle,fill,inner sep=1.5pt]{};
\node[right, align=left] at (axis cs:0,-1) {$x^1$};

\node at (axis cs:1,-0.6)[circle,fill,inner sep=1.5pt]{};
\node[right, align=left] at (axis cs:1,-0.5) {$x^2$};
\end{axis}
\end{tikzpicture}%
\caption{Three iterations of ART3. The feasible region is shaded while the constraints are dashed. The starting point $x^0$ is reflected into the first constraint to get to $x^1$, then into the second constraint to get to $x^2$. Since $x^2$ is in the feasible region, the algorithm terminates.}
\end{figure}

The algorithm ART3+O is an optimization algorithm that uses ART3+ as its workhorse. Given a linear optimization problem \eqref{eq:standardform} and an interval $[L,U]$ that is known to contain the optimal objective value, ART3+O uses bisection search to shrink the interval until it finds a $\varepsilon$-optimal value of \eqref{eq:standardform}. A description is given in Alg.~\ref{alg:art3plusO}. In each iteration, ART3+O determines the midpoint of the interval $M=(L+U)/2$ and then establishes whether \eqref{eq:standardform} has a feasible point with an objective value of at least $M$ by calling ART3+. If ART3+ finds such a feasible point, the search continues on the interval $[c^Tx,U]$, otherwise it continues on $[L,M]$, and repeats until the width of the interval is less than $\varepsilon$. The $x^{ART3+O}$ that is found, is at most $\varepsilon$ from the optimum, meaning that $c^Tx^{ART3+O} \geq c^Tx - \varepsilon$ for all $x \geq 0$ such that $Ax \leq b$.

A practical problem with ART3+O is that ART3+ runs indefinitely when there is no $x$ that satisfies $Ax \leq b$. Therefore, the algorithm needs to be stopped prematurely when it becomes unlikely that a feasible point will be found, e.g., after a large number of iterations, after a certain runtime, or when the iterates show a certain behavior \citep{de1985simultaneous,censor2003convergence}. This carries the risk of incorrectly concluding that an objective value of $M$ is unachievable. An $\varepsilon$-optimal solution can therefore not be guaranteed. In addition, the user is faced with a difficult trade-off: using a conservative stopping criterion means many cpu cycles are wasted on finding a solution that does not exist whereas a more aggressive criterion means that the final solution can be far from $\varepsilon$-optimal. What makes matters worse is that there is no guidance to what constitutes a ``conservative'' stopping criterion. As of today there is no known upper bound on the number of iterations ART3+ may take to find a feasible point.

This paper proposes an improvement to ART3+O so that it can accurately detect if an objective value of $M$ is attainable. It eliminates the need for an arbitrary stopping criterion, and guarantees an $\varepsilon$-optimal solution. With this improvement, the algorithm still terminates in a finite number of steps.

\section{Methods}
The cornerstone of the proposed method is Farkas' lemma. Although this lemma can be stated in many different ways, our results can be derived from the following form:
\begin{theorem} (Farkas' lemma) Exactly one of the following statements is true \citep{farkas2002}:
\begin{enumerate}
\item There exists an $x \in \mathbb{R}^n$ such that $Fx=b$ and $x \geq 0$.
\item There exists a $y \in \mathbb{R}^m$ such that $F^Ty \geq 0$ and $b^Ty \leq -1$.
\end{enumerate}
\end{theorem}
This lemma is a statement of alternatives. If the first statement is true, the second statement is not true and vice versa.

The algorithm ART3+O calls ART3+ to find a point $x$ in the set
\begin{align}\label{farkas:eq1}
\{x : c^Tx \geq M, Ax \leq b, x \geq 0\}.
\end{align}
Such a point exists if and only if there exists $s_1 \in \mathbb{R}$ and $s_2 \in \mathbb{R}^m$ such that:
\begin{align*}
\begin{pmatrix}-c^T & 1 & O \\ A & 0 & I\end{pmatrix} \begin{pmatrix}x \\ s_1 \\ s_2 \end{pmatrix} = \begin{pmatrix}-M \\ b\end{pmatrix}, \quad x \geq 0, \quad s_1 \geq 0, \quad s_2 \geq 0.
\end{align*}
We use these equations for the first statement of Farkas' lemma. The equations in the second statement are:
\begin{align}\label{farkas:eq2}
-c y_1 + A^T y_2 \geq 0, \quad -M y_1 + b^T y_2 \leq -1, \quad y_1 \geq 0, \quad y_2 \geq 0.
\end{align}
Therefore, the set \eqref{farkas:eq1} is empty if and only if there is a $y$ that satisfies \eqref{farkas:eq2}.

Currently, ART3+O cannot conclude with certainty that the set \eqref{farkas:eq1} is empty. It presumes that the set is empty when ART3+ cannot find a point in that set within the allocated runtime or number of iterations, but that presumption may be wrong.

\begin{algorithm}\label{alg:art3plusOimproved}
	\KwData{$A \in \mathbb{R}^{m \times n}$, $b \in \mathbb{R}^{m}$, $c \in \mathbb{R}^n$, $L \in \mathbb{R}$, $U \in \mathbb{R}$}
	\KwResult{$x$ that solves \eqref{eq:standardform}}
	\While{$U-L > \varepsilon$}{
		$(L+U)/2 \to M$\;
		Call two instances of ART3+ in parallel:
		\begin{enumerate}
		\item to find $x$ such that $\begin{pmatrix}-c^T \\ A \\ -I\end{pmatrix} x \leq \begin{pmatrix} -M \\ b \\ 0\end{pmatrix}$
		\item to find $y$ such that $\begin{pmatrix}c & -A^T \\ -M & b^T \\ -1 & 0 \\ O & -I\end{pmatrix} \begin{pmatrix}
		y_1 \\ y_2 \end{pmatrix} \leq \begin{pmatrix} 0 \\ -1 \\ 0 \\ 0\end{pmatrix}$\;
        \end{enumerate}
        When one instance of ART3+ finds a solution, terminate the other instance\;
		\eIf{a feasible $x$ was found}{$c^Tx \to L$}{$M \to U$}
	}
	\caption{The suggested improved algorithm.}
\end{algorithm}

We propose to run ART3+ both on the set \eqref{farkas:eq1} and on the inequalities \eqref{farkas:eq2} in parallel, until it finds a point in either set. When it finds a point that satisfies \eqref{farkas:eq2}, that is mathematical proof that the set \eqref{farkas:eq1} is empty and that an objective value of $M$ is unachievable. Therefore, bisection search can continue on the interval $[L,M]$. This algorithm is outlined in Alg.~\ref{alg:art3plusOimproved}.

ART3+O requires that the rows of the constraint matrix $A$ can be accessed efficiently, while the improved algorithm also requires efficient access to the columns of $A$. If only a single copy of $A$ is kept in memory, one of the two instances of ART3+ cannot have sequential data access, causing unnecessary slowness. In radiation therapy, this is an issue because the matrix $D$ in \eqref{eq:dose} is often stored in sparse format, meaning that only the indices and values of nonzero elements get stored. For sparse formats, either the rows or the columns are efficiently accessible, but not both. Therefore, the $D$ matrix needs to be transposed. Consequently, the memory requirements of the improved algorithm are roughly twice as high as those of ART3+O. The difference in runtime is harder to predict. Although Alg.~\ref{alg:art3plusOimproved} runs two threads in parallel and therefore uses twice as much cpu power when a feasible point $x$ is found, it is possible that the second thread terminates early enough when no feasible $x$ exists to make up for this.

Since ART3+ terminates in a finite number of steps if the feasible set is full dimensional \citep{herman2008fast}, it trivially follows that Alg.~\ref{alg:art3plusOimproved} also terminates in a finite number of steps. The requirement of full dimensionality is not an actual restriction: techniques from the ellipsoid method can be used to slightly perturb the inequalities in a way that the feasible region becomes full dimensional if it is not empty, while it remains empty if the original feasible region is empty \citep{bland1981ellipsoid}. Nevertheless, we shall validate the full dimensionality assumption in our numerical results by subtracting a small number ($10^{-3}$) from the right-hand side of the inequalities given to ART3+ by Alg.~\ref{alg:art3plusOimproved} and show that a point satisfying those inequalities exists.

We have taken the implementation of ART3+O from the in-house treatment planning system Astroid and modified it to make it accurately correspond to Alg.~\ref{alg:art3plusO}, with a stopping criterion of $2\cdot 10^7$ iterations and $\varepsilon=0.1$ \citep{chen2010fast}. To implement Alg.~\ref{alg:art3plusOimproved}, we added multithreading capabilities and code to transpose a sparse matrix. As the starting point for ART3+, we took the zero vector in the first call, and the final point from the previous call in subsequent calls. This makes the implementation nondeterministic, because the thread that gets terminated without finding a point has run a nondeterministic number of iterations.

The method is tested on a small radiation therapy case with 227 pencil beams. The objective is to maximize the minimum dose in the 256 voxel PTV, while the maximum dose in a surrounding structure of 8406 voxels is constrained to 50 Gy. The relevant part of the $D$ matrix has 288,610 nonzero elements ($\approx 15\%$). The auxiliary variable necessary to model a minimum dose objective in the format of \eqref{eq:standardform} was replaced with $M$ to avoid unnecessary overhead. So, the optimization problem is $\max_{x \geq 0,t \geq 0}\{t : D_1 x \geq t, D_2 x \leq 50e\}$, but the inequalities given to ART3+ in Alg.~\ref{alg:art3plusOimproved} are $\{D_1 x \geq Me, D_2 x \leq 50e, x \geq 0\}$ and $\{-D_1 y_{11} + D_2 y_{12} \leq 0, M e^T y_{11} - 50 e^T y_{12} \leq -1, y \geq 0 \}$.

The method is also tested on a larger case with 5384 pencil beams and approximately $2 \cdot 10^6$ constraints that limit the mean, minimum or maximum dose, and an objective to minimize the maximum dose in a structure with $3 \cdot 10^5$ voxels. The objective was treated in the same way as the small case.

The method is tested on another 77 clinical test cases spanning 11 patients with different tumor sites. Each patient has a common set of constraints, but different objectives (at most 10 per patient). In addition to minimum and maximum dose objectives and constraints, some cases have mean dose constraints or a mean underdose objective. The $D$ matrices vary in size between 12 MB and 2.5 GB. These cases will be run on the ERISone cluster (Partners Healthcare, Boston, MA, USA), which uses Intel E5-2690 v3 and Intel X5670 CPUs,  with a time limit of 1 day of cpu-time (around 12 hours of wall-clock time).

The first two cases are run on a machine with two Intel E5-2687W v3 cpus. CPLEX 12.8, a state of the art solver for linear optimization, is used to obtain the optimal solution for both cases and to validate the full dimensionality assumption. Although the second case is large and CPLEX runs out of memory if it attempts to solve it, it can solve an equivalent reformulation. The case has one $D$ matrix per beam, $D^1$ and $D^2$, with corresponding $x^1$ and $x^2$, and constraints are formulated on $D^1x^1$, on $D^2x^2$, and on $D^1 x^1 + D^2x^2$. By adding auxiliary variables $d^1 = D^1 x^1$ and $d^2 = D^2 x^2$, we avoid that rows of $D^k$ get duplicated across constraints, which resolves the memory issue. To resolve the memory issue when trying to validate the full dimensionality assumption for the $y$-space, we take the feasibility problem for $x$ and turn it into an optimization problem by adding the objective of maximizing $-\sum x_i$. The dual variables of the inequality constraints then satisfy $c y_1 - A^T y_2 \leq -1$, $b^Ty<0$, and $y\geq 0$ by LP duality. We then slightly increase each element of $y$ such that $y>0$.

\section{Results}
CPLEX solves the small case in 0.8 seconds and shows that the optimal value is 32.71 Gy. We confirmed that the feasible set is full dimensional in the first instance of ART3+ as long as $M$ is less than the optimal value, while it is full dimensional in the second instance of ART3+ when $M$ is larger than the optimal value. ART3+O takes 12.5 seconds to obtain a solution with value 32.57. Table \ref{tbl:smallcase} shows the steps taken by Alg.~\ref{alg:art3plusOimproved} during its runtime of around 2.5 hours. In steps 2, 4, 5 and 9, it was proven that the objective values 37.51, 34.38, 32.82 and 32.77 are unachievable, while in steps 1, 3, 6, 7 and 8 the objective value was gradually improved to 32.71. Although this final solution is guaranteed to be within 0.1 Gy of the optimum, it is in fact only 0.0015 Gy from the optimum. Looking at the number of iterations, there are two remarkable observations. The first is that it takes a considerable number of iterations to conclude that a value of $M$ is unattainable. Even for $M=37.5$ (Step 2), which is far from achievable, $25 \cdot 10^6$ iterations are necessary. The second is that close to optimality, even the regular ART3+ may take a large number of iterations. In Step 8, $29\cdot 10^9$ iterations are needed to obtain a solution for $M=32.71$. This is orders of magnitude higher than the iteration limit of $2\cdot 10^7$ used by \citet{chen2010fast}.

\begin{table}
	\centering
	\begin{tabular}{llllrr}
		\toprule
		Step & $L$ & $U$ & Farkas & Time (s) & Iterations \\
		\midrule
		1    &  0.00 & 50.00 & 1 & 0.1  & 62,356 \\
		2    & 25.01 & 50.00 & 2 & 1.9 & 24,948,327 \\
		3    & 25.01 & 37.51 & 1 & 0.0  & 332,498 \\
		4    & 31.26 & 37.51 & 2 & 11.1 & 147,789,618 \\
		5    & 31.26 & 34.38 & 2 & 3160.4 & 39,374,210,616 \\
		6    & 31.26 & 32.82 & 1 & 0.0  & 53,662 \\
		7    & 32.38 & 32.82 & 1 & 0.1  & 2,477,787 \\
		8    & 32.60 & 32.82 & 1 & 1345.9 & 29,435,448,946 \\
		9    & 32.71 & 32.82 & 2 & 4618.7 & 57,541,614,242 \\
		\bottomrule
	\end{tabular}
	\caption{Results of Alg.~\ref{alg:art3plusOimproved} on the small radiation therapy case. $[L,U]$ is the search interval of bisection search, Farkas indicates whether the first or second statement in Farkas' lemma was found to be true, and the last two columns show the runtime statistics of the instance of ART3+ that found a solution.}
	\label{tbl:smallcase}
\end{table}

CPLEX (barrier) solves the large case in just under 3.5 hours with an optimal value of 50.8 Gy. ART3+O finds a value of 51.6 Gy in 847.4 seconds. Alg.~\ref{alg:art3plusOimproved} takes 8 seconds to transpose the $D$ matrix. In the two steps of bisection search, the second instance of ART3+ proves that $M=26.47$ and $M=39.70$ are unattainable in approximately 6 minutes and 6 hours, respectively. We terminated bisection search in the third step of bisection search, because even after 100 hours and $218 \cdot 10^9$ iterations, ART3+ could not prove that $M=46.32$ is unattainable, despite that the full dimensionality assumption holds for this value of $M$.

For the test set of 77 cases, ART3+O does not find a feasible point within the iteration limit for half the patients, affecting 43 of the optimization runs, and incorrectly concludes that the problem is infeasible. We therefore re-ran ART3+O with an iteration limit of $10^9$ instead of $2\cdot 10^7$, affecting the runtimes by a median factor of 26. Still, 2 out of 10 patients are incorrectly classified as infeasible, affecting 13 optimization runs. The remaining 64 runs took between 8.5 seconds and 3:05 hours (median: 35 minutes). The suboptimality is between 0 and 3.6 Gy (median: 0.2 Gy), and is less than 0.1 Gy in 13 out of 64 cases. Alg.~\ref{alg:art3plusOimproved} finds a feasible point in all cases. For the 64 cases that ART3+O could solve, Alg.~\ref{alg:art3plusOimproved} finishes before the time limit in 7 cases, but is always slower than ART3+O. The suboptimality ranges between 0 Gy and 12.5 Gy (median: 1.1 Gy). Alg.~\ref{alg:art3plusOimproved} outperforms ART3+O by at least 0.1 Gy in 14 cases, while the opposite is true in 41 cases. Alg.~\ref{alg:art3plusOimproved} solves the 13 cases that ART3+O incorrectly declares infeasible, with the suboptimality ranging between 0 and 5.7 Gy (median 0.2 Gy).

\section{Discussion}
The results presented by \citet{chen2010fast} show that ART3+O can be used to quickly optimize a treatment plan. Its main weakness is the arbitrary iteration limit. In our numerical examples, we found out that this weakness is not purely theoretical, but that it has practical impact. The iteration limit suggested by \citet{chen2010fast} is too low, and even using a considerably higher number still leads to suboptimalities of up to 3.6 Gy. The theoretical advantage of the proposed alternative does not translate into a practical advantage. Despite the significantly higher cpu time (24 hours versus 3:05 hours), the final objective value is often worse.

The dual problem of \eqref{eq:standardform} is $\min\{b^Ty : A^T y \geq c, y \geq 0\}$. Alternative solvers for linear optimization, notably the simplex method and an interior point method, generate not just an optimal solution $x^*$ to \eqref{eq:standardform}, but also an optimal dual solution $y^*$ as a byproduct. Due to strong duality, $c^Tx^* = b^Ty^*$ if the problem is feasible and bounded. For $M\leq c^Tx^*$, $x^*$ is a solution for the first instance of ART3+, while for $M>c^Tx^*$, $y_1 = 1/(M-b^Ty^*)$ and $y_2 = y_1 y^*$ is a solution for the second instance of ART3+ in Alg.~\ref{alg:art3plusOimproved}. That means that alternative solvers solve the same two problems as Alg.~\ref{alg:art3plusOimproved}.

It is an open question as to why ART3+ often finds near optimal solutions in a moderate number of iterations, while it takes significantly more iterations to prove that a certain objective value is unattainable. A possible explanation is that although the primal can have a large number of constraints, many are correlated or redundant, such as constraints on the maximum dose in two neighboring voxels. Correlated constraints do probably not occur in the second instance of ART3+ in Alg.~\ref{alg:art3plusOimproved}, because that would require two different $x_j$ to deliver almost the same dose. The phenomenon that it is easy to find an $x$ and hard to find a $y$ is problem specific. Outside of the domain of radiation therapy there are problems for which the opposite is true. An artificial example can be constructed by taking the dual of \eqref{eq:standardform} for a radiation therapy problem.

Our improvement to ART3+O can be applied to ART3+ as well. ART3+ was originally proposed for image reconstruction by \citet{herman1975relaxation} where it approximately solves a large system $Ax=b$ by finding a solution to
\begin{align}\label{eq:art3}b-\varepsilon \leq Ax \leq b + \varepsilon.\end{align}
\citet{herman1975relaxation} suggested to select $\varepsilon$ based on prior knowledge or experimentally, but there is a delicate balance. When $\varepsilon$ is chosen too small, there may be no solution, while a large value allows for solutions that do not accurately reconstruct the original distribution. It is therefore desirable to determine the smallest $\varepsilon$ for which a solution exists. By Farkas' lemma, there is either an $x$ that satisfies \eqref{eq:art3}, or a $y$ that satisfies $A^T y_1 - A^T y_2 \geq 0$, $(b+\varepsilon)^Ty_1 - (b-\varepsilon)^Ty_2 \leq -1$, $y_1 \geq 0$, $y_2 \geq 0$. One could therefore run ART3+ in parallel on both sets of inequalities to establish if there is a solution to \eqref{eq:art3} for a given value of $\varepsilon$.

It is well known that the order in which the projections are performed has a significant impact on the performance of a projection algorithm. Ideally, consecutive constraints are as orthogonal as possible \citep{guan1994projection}. To test the impact on our results, we have reordered the constraints in a way that consecutive constraints are as orthogonal as possible. We did this greedily, starting with the nonnegativity constraints (which are fully orthogonal), and then adding the constraints one by one such that the added constraint was as orthogonal as possible to the previous one. Due to the computational cost of this ordering heuristic, we only tested it on the small case. To our surprise, the number of iterations {\it in}creased for the second instance of ART3+, making the overall performance worse. An alternative ordering based on the golden ratio \citep{kohler2004projection} did not improve the performance either.

The lack of a good stopping criterion has also been noted for projection methods in {\it convex} optimization \citep{gibali2014generalized}. The conic variant of Farkas' lemma could be used for such problems. Let $f_i$ be closed and convex functions and let $f_i^*$ denote their convex conjugate functions. Exactly one of the following sets is not empty:
\begin{align}
&1. \; \{ x \in \mathbb{R}^n : f_i(x) < 0 \quad i=1,\ldots,m\} \label{convexset1} \\
&2. \; \{ \lambda \in \mathbb{R}^m_+, v^i \in \mathbb{R}^n : \sum_{i=1}^m v^i = 0, \sum_i \lambda_i f^*\left(\frac{v^i}{\lambda_i}\right) \leq -1 \}. \label{convexset2}
\end{align}
The derivation is given in Appendix \ref{apdx:convex}. A necessary step toward full dimensionality is substituting out one of the vectors $v^i$. By using the two sets \eqref{convexset1} and \eqref{convexset2}, the methods in this paper extend to convex optimization.
\section{Conclusion}
ART3+O cannot guarantee $\varepsilon$-optimality of the final solution due to an arbitrary stopping criterion. An improvement that provides such a guarantee was suggested, but despite its theoretical basis, its runtime is too high, especially in comparison with alternative solvers. If an optimality guarantee is not required, ART3+O remains a viable algorithm, but their users should be aware that a solution may be suboptimal.
\appendix
\section{Extension to convex optimization}\label{apdx:convex}
\begin{theorem} (Conic variant of Farkas' lemma) Let $K$ be a closed convex cone with dual cone $K^*$ and let $F \in \mathbb{R}^{m \times n}$. Exactly one of the following statements is true \citep{craven1977generalizations}:
\begin{enumerate}
\item There exists an $x \in \mathbb{R}^n$ such that $Fx=b$ and $x \in K$.
\item There exists a $y \in \mathbb{R}^m$ such that $F^Ty \in K^*$ and $b^Ty \leq -1$.
\end{enumerate}
\end{theorem}
The closure of the set \eqref{convexset1} can be denoted as $\{(x,t) : t=1, (x,t) \in K = \cap_{i=1}^m K_i\}$, with $K_i = \{(x,t) : t f_i(x/t) \leq 0, t \geq 0\}$ and $0 f_i(x/0) = \lim_{t \downarrow 0} tf_i(x/t)$. Note that $K$ is indeed a closed convex cone. The $x$ in Farkas' lemma is actually the vector $(x,t) \in \mathbb{R}^{n+1}$, $F=(0,\ldots,0,1) \in \mathbb{R}^{1 \times (n+1)}$, and the second set in Farkas' lemma is $\{y \in \mathbb{R} : (0,\ldots,0,y) \in K^*, y \leq -1\}$. The condition $(0,\ldots,0,y) \in K^*$ is by definition equivalent to $ty \geq 0$ $\forall (x,t) \in \cap_{i=1}^m K_i$. To eliminate the ``$\forall$'' quantifier, we rewrite this definition as $\min_{x,t \geq 0} \{ty : t f_i(x/t) \leq 0 \; i=1,\ldots,m\} \geq 0$, and then apply Lagrange duality to the minimization problem using the method by \citet{roos2018}. The objective function of this problem is $g_0(x,t) = ty$ and the constraint functions are $g_i(x,t) = t f_i (x/t)$. The corresponding conjugate functions are $g_0^*(v^0,u^0) = 0$ if $v^0=0$ and $u^0=y$ ($\infty$ otherwise) and $g_i^*(v^i,u^i) = 0$ if $f^*(v^i) + u^i \leq 0$ ($\infty$ otherwise). Therefore, by \citet{roos2018}, the Lagrange dual is $\max_{\lambda \geq 0,v,u}\{ 0 : \sum_{i=1}^m v^i = 0, y+\sum_{i=1}^m u^i = 0, \lambda_i f^*(v^i / \lambda_i) + u^i \leq 0 \}$. The second statement in Farkas lemma reduces to: there exists a $y \in \mathbb{R}$, $\lambda \in \mathbb{R}^m_+$, $v^i \in \mathbb{R}^n$ and $u^i \in \mathbb{R}$ such that $\sum_{i=1}^m v^i = 0$, $y+\sum_{i=1}^m u^i = 0$, $\lambda_i f^*(v^i / \lambda_i) + u^i \leq 0$ and $y \leq -1$. Substituting out $y$ and $\sum_i u_i$, we get the set \eqref{convexset2}.

\section*{Acknowledgment}
Supported in part by NIH U19 Grant 5U19CA021239-38.

\bibliographystyle{abbrvnatnew}
\bibliography{bibfile}

\begin{thebibliography}{22}
\providecommand{\natexlab}[1]{#1}
\providecommand{\url}[1]{\texttt{#1}}
\expandafter\ifx\csname urlstyle\endcsname\relax
  \providecommand{\doi}[1]{doi: #1}\else
  \providecommand{\doi}{doi: \begingroup \urlstyle{rm}\Url}\fi

\bibitem[Bahr et~al.(1968)Bahr, Kereiakes, Horwitz, Finney, Galvin, and
  Goode]{bahr1968method}
G.~Bahr, J.~Kereiakes, H.~Horwitz, R.~Finney, J.~Galvin, and K.~Goode.
\newblock The method of linear programming applied to radiation treatment
  planning.
\newblock \emph{Radiology}, 91\penalty0 (4):\penalty0 686--693, 1968.

\bibitem[Bauschke and Borwein(1996)]{bauschke1996projection}
H.~H. Bauschke and J.~M. Borwein.
\newblock On projection algorithms for solving convex feasibility problems.
\newblock \emph{SIAM review}, 38\penalty0 (3):\penalty0 367--426, 1996.

\bibitem[Bland et~al.(1981)Bland, Goldfarb, and Todd]{bland1981ellipsoid}
R.~G. Bland, D.~Goldfarb, and M.~J. Todd.
\newblock The ellipsoid method: A survey.
\newblock \emph{Operations Research}, 29\penalty0 (6):\penalty0 1039--1091,
  1981.

\bibitem[Bortfeld et~al.(1990)Bortfeld, B{\"u}rkelbach, Boesecke, and
  Schlegel]{bortfeld1990methods}
T.~Bortfeld, J.~B{\"u}rkelbach, R.~Boesecke, and W.~Schlegel.
\newblock Methods of image reconstruction from projections applied to
  conformation radiotherapy.
\newblock \emph{Physics in Medicine \& Biology}, 35\penalty0 (10):\penalty0
  1423, 1990.

\bibitem[Breedveld et~al.(2017)Breedveld, van~den Berg, and
  Heijmen]{Breedveld2016}
S.~Breedveld, B.~van~den Berg, and B.~Heijmen.
\newblock An interior-point implementation developed and tuned for radiation
  therapy treatment planning.
\newblock \emph{Computational Optimization and Applications}, 68\penalty0
  (2):\penalty0 209--242, 2017.

\bibitem[Censor and Tom(2003)]{censor2003convergence}
Y.~Censor and E.~Tom.
\newblock Convergence of string-averaging projection schemes for inconsistent
  convex feasibility problems.
\newblock \emph{Optimization Methods and Software}, 18\penalty0 (5):\penalty0
  543--554, 2003.

\bibitem[Chen et~al.(2010)Chen, Craft, Madden, Zhang, Kooy, and
  Herman]{chen2010fast}
W.~Chen, D.~Craft, T.~M. Madden, K.~Zhang, H.~M. Kooy, and G.~T. Herman.
\newblock A fast optimization algorithm for multicriteria intensity modulated
  proton therapy planning.
\newblock \emph{Medical physics}, 37\penalty0 (9):\penalty0 4938--4945, 2010.

\bibitem[Chen et~al.(2012)Chen, Unkelbach, Trofimov, Madden, Kooy, Bortfeld,
  and Craft]{chen2012including}
W.~Chen, J.~Unkelbach, A.~Trofimov, T.~Madden, H.~Kooy, T.~Bortfeld, and
  D.~Craft.
\newblock Including robustness in multi-criteria optimization for
  intensity-modulated proton therapy.
\newblock \emph{Physics in Medicine \& Biology}, 57\penalty0 (3):\penalty0
  591--608, 2012.

\bibitem[Craven and Koliha(1977)]{craven1977generalizations}
B.~D. Craven and J.~J. Koliha.
\newblock Generalizations of {F}arkas' theorem.
\newblock \emph{SIAM Journal on Mathematical Analysis}, 8\penalty0
  (6):\penalty0 983--997, 1977.

\bibitem[De~Pierro and Iusem(1985)]{de1985simultaneous}
A.~R. De~Pierro and A.~N. Iusem.
\newblock A simultaneous projections method for linear inequalities.
\newblock \emph{Linear Algebra and its applications}, 64:\penalty0 243--253,
  1985.

\bibitem[Farkas(1902)]{farkas2002}
J.~Farkas.
\newblock \"{U}ber die {T}heorie der {E}infachen {U}ngleichungen.
\newblock \emph{Journal f\"{u}r die Reine und Angewandte Mathematik},
  124:\penalty0 1--27, 1902.

\bibitem[Gibali et~al.(2014)Gibali, K{\"u}fer, Reem, and
  S{\"u}ss]{gibali2014generalized}
A.~Gibali, K.-H. K{\"u}fer, D.~Reem, and P.~S{\"u}ss.
\newblock A generalized projection-based scheme for solving convex constrained
  optimization problems.
\newblock \emph{Computational Optimization and Applications}, 1--26, 2014.

\bibitem[Guan and Gordon(1994)]{guan1994projection}
H.~Guan and R.~Gordon.
\newblock A projection access order for speedy convergence of {ART} (algebraic
  reconstruction technique): a multilevel scheme for computed tomography.
\newblock \emph{Physics in Medicine \& Biology}, 39\penalty0 (11):\penalty0
  2005, 1994.

\bibitem[Herman(1975)]{herman1975relaxation}
G.~T. Herman.
\newblock A relaxation method for reconstructing objects from noisy x-rays.
\newblock \emph{Mathematical Programming}, 8\penalty0 (1):\penalty0 1--19,
  1975.

\bibitem[Herman and Chen(2008)]{herman2008fast}
G.~T. Herman and W.~Chen.
\newblock A fast algorithm for solving a linear feasibility problem with
  application to intensity-modulated radiation therapy.
\newblock \emph{Linear algebra and its applications}, 428\penalty0
  (5-6):\penalty0 1207--1217, 2008.

\bibitem[Kohler(2004)]{kohler2004projection}
T.~Kohler.
\newblock A projection access scheme for iterative reconstruction based on the
  golden section.
\newblock In \emph{Nuclear Science Symposium Conference Record, 2004 IEEE},
  volume~6, 3961--3965, 2004.

\bibitem[Lessard and Pouliot(2001)]{lessard2001inverse}
E.~Lessard and J.~Pouliot.
\newblock Inverse planning anatomy-based dose optimization for
  {HDR}-brachytherapy of the prostate using fast simulated annealing algorithm
  and dedicated objective function.
\newblock \emph{Medical Physics}, 28\penalty0 (5):\penalty0 773--779, 2001.

\bibitem[Lomax(1999)]{lomax1999intensity}
A.~Lomax.
\newblock Intensity modulation methods for proton radiotherapy.
\newblock \emph{Physics in Medicine \& Biology}, 44\penalty0 (1):\penalty0 185,
  1999.

\bibitem[Romeijn et~al.(2003)Romeijn, Ahuja, Dempsey, Kumar, and
  Li]{romeijn2003novel}
H.~E. Romeijn, R.~K. Ahuja, J.~F. Dempsey, A.~Kumar, and J.~G. Li.
\newblock A novel linear programming approach to fluence map optimization for
  intensity modulated radiation therapy treatment planning.
\newblock \emph{Physics in Medicine \& Biology}, 48\penalty0 (21):\penalty0
  3521--3542, 2003.

\bibitem[Roos et~al.(2018)Roos, Balvert, Gorissen, and den Hertog]{roos2018}
C.~Roos, M.~Balvert, B.~L. Gorissen, and D.~den Hertog.
\newblock A universal and structured way to derive dual optimization problem
  formulations.
\newblock \emph{Optimization Online}, 2018.

\bibitem[Shepard et~al.(1999)Shepard, Ferris, Olivera, and
  Mackie]{shepard1999optimizing}
D.~M. Shepard, M.~C. Ferris, G.~H. Olivera, and T.~R. Mackie.
\newblock Optimizing the delivery of radiation therapy to cancer patients.
\newblock \emph{Siam Review}, 41\penalty0 (4):\penalty0 721--744, 1999.

\bibitem[Trofimov et~al.(2012)Trofimov, Unkelbach, and
  Craft]{trofimov2012treatment}
A.~Trofimov, J.~Unkelbach, and D.~Craft.
\newblock Treatment-planning optimization.
\newblock In H.~Paganetti, editor, \emph{Proton therapy physics}, 461--484. CRC
  Press/Taylor \& Francis, Boca Raton, FL, USA, 2012.

\end{thebibliography}
\end{document}